# Tipping Points, Pulse Elasticity and Tonal Tension: An Empirical Study on What Generates Tipping Points

**Canishk Naik**
Department of Mathematics
London School of Economics
canishknaik@gmail.com

**Elaine Chew**
Centre for Digital Music
School of Electronic Eng and Comp Science
Queen Mary University of London
elaine.chew@qmul.ac.uk

## ABSTRACT

Tipping points are moments of change that characterise crucial turning points in a piece of music. This study presents a first step towards quantitatively and systematically describing the musical properties of tipping points. Timing information and computationally-derived tonal tension values which correspond to dissonance, distance from key, and harmonic motion are compared to tipping points in Ashkenazy's recordings of six Chopin Mazurkas, as identified by 35 listeners. The analysis shows that all popular tipping points but one could be explained by statistically significant timing deviations or changepoints in at least one of the three tension parameters.

## 1. MOTIVATION

The term tipping points was first introduced in [1] in reference to temporal deviations that simulate the physical sensation of a tipping point. This is implemented in performance through significant slowing of tempo. Temporal tipping points are also associated with apices of musical tension that dissipate when the performer releases the subsequent notes. So, not all slowings of tempo are tipping points, and not all tipping points are created through temporal deviations. Listeners can often pinpoint tipping points in musical performance. However, it can be hard to rigorously define the characteristics of a tipping point. This serves as the motivation for the current study, which investigates whether computational methods can be used to illuminate what makes a tipping point a tipping point.

The method we employ compares listener-defined tipping points to musical features that are frequently associated with tipping points, namely, timing deviations and musical tension. The interplay between pulse elasticity and tonal tension has been discussed in both [1] and [2]. In [1], temporal tipping points are described as moments where information is peaked and what is to come is known; tension then emerges from the listener not knowing *when* the certain outcome will take place. The tipping points were annotated by the author, but not checked against a computational model of tension. In [2], Gingras et al. used IDyOM to model tension using information theoretic concepts, focusing on melodic expectation and perceived tension. [2] considered how changes in information content affected performance timing and found that performers tend to slow down when anticipating unexpected notes.

Our objective is to compare tipping points located by computational models —both time-based and based on a computational model of tonal tension—with those perceived by listeners so as to systematically determine the properties of tipping points. This represents the first empirical study on tipping points. We further explore computationally the interplay of tonal tension and pulse elasticity.

## 2. METHODOLOGY

In order to ascertain where people perceive tipping points, we create a short questionnaire comprising of six test recordings, each an excerpt (less than two minutes long) from performances of Chopin Mazurkas by Vladimir Ashkenazy: Op 6 Nr 4, Op 17 Nr 2, Op 30 Nr 1, Op 33 Nr 1, Op 33 Nr 3 and Op 41 Nr 3. A short tutorial was created to guide participants through two illustrative tipping point examples; listeners are then asked to indicate one tipping point in each recording. We also ask how far in advance listeners anticipated the tipping point and how tense the tipping point felt. In this experiment, we view the participant-annotated tipping points as ground truth and aim to predict their selections using two computational models.

**Model 1:** This is a time-based model that predicts tipping points when the beat duration is larger than a specified threshold. Defining $d_i$ as the duration of beat $i$ in performance time, we write a program in MATLAB to indicate that there is a time-based tipping point at the time where the $i^{\text{th}}$ beat starts if:

$$d_i > \bar{d} + 2.5 \times \sigma, \qquad (1)$$

where $\bar{d}$ is the sample mean of the $\{d_i\}$ and $\sigma$ is the sample standard deviation. The program rejects any tipping points which are found in the first and last three bars of a piece since performers often play at a slower pace in the introductory and concluding passages.

**Model 2:** This is a tonal tension-based model that predicts tipping points as statistically significant changes in the mean or variance of the tonal tension data. Tonal tension is estimated through the spiral array-based computational model described in [3], which captures aspects of dissonance (cloud diameter), harmonic change (cloud momentum), and distance from key (tensile strain). We identify statistically significant change using the 'changepoint' package [4] in R. We are especially interested in points where changes represent shifts from high to low tension. We take a long build-up and quick release of tension as an indication of a tipping point.

## 3. RESULTS

First, the listener-identified tipping points showed a wide spread, and had an across-Mazurka average standard deviation of 15.9 seconds.

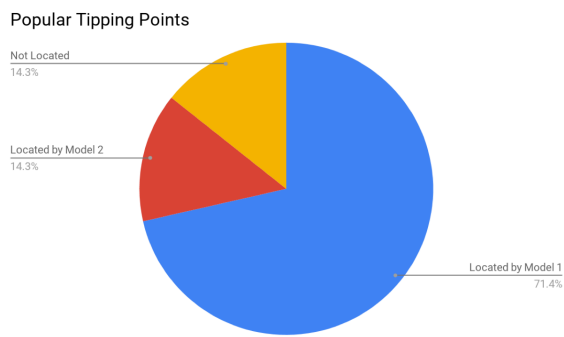

**Figure 1**. Proportion of perceived tipping points located by Models 1&2.

A tipping point is considered popular when at least 25% of the listeners perceived a tipping point within a symmetric 4s interval around that point. From Figure 1 we can see almost all popular tipping points could be explained as statistically significant timing deviations or changepoints in a tension parameter. The one exception was a tipping point marked by 26% of participants in Op 33 Nr 1, between 41-45s. Neither Model 1 nor Model 2 located this point. We discuss this in Section 4.

**Case Example:** In Mazurka Op 17 Nr 2, 40% of participants perceived a tipping point between 38-42 seconds, as shown in the top graph in Fig 2. Model 1 located a tipping point at 39s (see bottom graph in Fig 2). Model 1 also located a tipping point at 52 seconds; only 5% of listeners chose this. 23% perceived a tipping point between 75-83s. This was not identified by Model 1; however, Model 2 detected a changepoint at 77s from high to low key distance, as shown in Fig. 3.

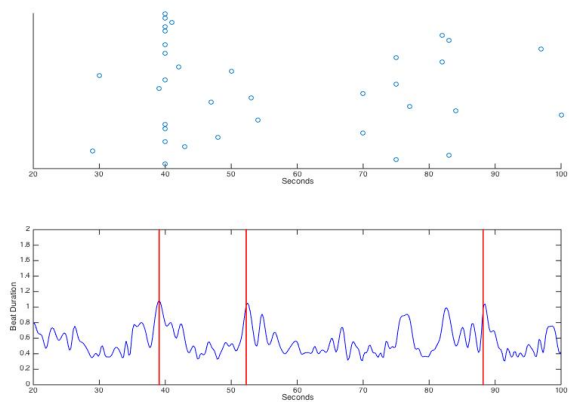

**Figure 2**. Scatter plot of perceived tipping points (top); tipping points predicted by Model 1 (bottom)

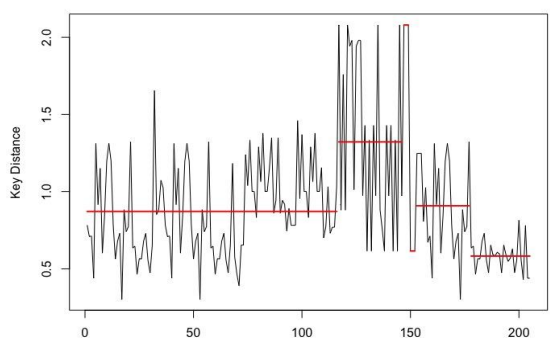

**Figure 3**. Changepoints in the key distance parameter.

## 4. DISCUSSION AND CONCLUSIONS

As we expected, listeners often reported perceived tipping points at critical turning points in the music. These commonly fell at high-tension structural boundaries in the pieces, before release and re-emergence of the theme, for example at 75-83s in Op 17 Nr 2.

However, participants may have struggled to wholly grasp the concept. Sometimes, listeners perceived tipping points at structural boundaries which marked the ends of sections in a final and stable way. For example, between 56-60s in Op 33 Nr 1, 57% of participants perceived a tipping point and this was verified by Model 1. 41-45s in Op 33 Nr 1 was also an example and here, neither model verified the response. We do not consider these as tipping points since there is insufficient tension.g point.

To improve the analyses and take the exploration further, there are several steps we may take. Future versions of the questionnaire could allow participants the freedom to decide the number of tipping points instead of constraining them to choosing only one. Information theoretic techniques to measure expectation can be incorporated as an alternative to, or in addition to, tonal tension measures as predictors of perceived tension. For example, delay of harmonic closure in a perfect cadence is particularly potent as a device for inducing tension and this would be identified using expectation measures. We are also keen to analyse computationally which musical features elicit strong emotions during a tipping point.